\documentclass{sig-alternate}
\usepackage{amsmath,amsfonts,dsfont}
\usepackage{multirow}
\usepackage{paralist}
\usepackage{graphicx}
\usepackage{enumitem}
\usepackage{booktabs}
\usepackage{mymacros}
\usepackage[numbers]{natbib}
\usepackage{algpseudocode}
\usepackage{algorithm}
\usepackage{bm}

\setcitestyle{square,citesep={;},aysep={},yysep={;}}

\setcopyright{acmcopyright}

\begin{document}
\bibliographystyle{agsm_nourl}

\hyphenation{bay-es-i-an}
\hyphenation{mar-g-i-n-al}
\title{Continuous Monitoring of A/B Tests without Pain: Optional Stopping in Bayesian Testing }

\numberofauthors{3} 
\author{
\alignauthor
Alex Deng  \thanks{First and second authors contributed equally to this work.}\\ 
      \affaddr{Microsoft}\\
       \affaddr{One Microsoft Way}\\
       \affaddr{Redmond, WA 98052}\\
       \email{alexdeng@microsoft.com}
\alignauthor
Jiannan Lu \textsuperscript{*}\\
      \affaddr{Microsoft}\\
       \affaddr{One Microsoft Way}\\
       \affaddr{Redmond, WA 98052}\\
       \email{jiannl@microsoft.com}
\alignauthor
Shouyuan Chen\\
      \affaddr{Microsoft}\\
       \affaddr{One Microsoft Way}\\
       \affaddr{Redmond, WA 98052}\\
       \email{shouch@microsoft.com}
}  
\date{1 Jan 2015}

\maketitle
\begin{abstract}
A/B testing is one of the most successful applications of statistical theory in modern Internet age. One problem of Null Hypothesis Statistical Testing (NHST), the backbone of A/B testing methodology, is that experimenters are not allowed to continuously monitor the result and make decision in real time. Many people see this restriction as a setback against the trend in the technology toward real time data analytics. Recently, Bayesian Hypothesis Testing, which intuitively is more suitable for real time decision making, attracted growing interest as an alternative to NHST. While corrections of NHST for the continuous monitoring setting are well established in the existing literature and known in A/B testing community, the debate over the issue of whether continuous monitoring is a proper practice in Bayesian testing exists among both academic researchers and general practitioners. In this paper, we formally prove the validity of Bayesian testing with continuous monitoring when proper stopping rules are used, and illustrate the theoretical results with concrete simulation illustrations. We point out common bad practices where stopping rules are not proper and also compare our methodology to NHST corrections. General guidelines for researchers and practitioners are also provided. 
\end{abstract}

\vspace{1mm}
\noindent
{\bf Category and Subject Descriptors:} G.3 {[Probability and Statistics]}: {Statistical Computing}

\vspace{1mm}

\section{Introduction}

Many online service companies nowadays have been using online controlled experiments, a.k.a. A/B Testing, as a scientifically grounded way to evaluate changes and comparing different alternatives. A/B testing plays a leading role in establishing the mantra of data driven decision making, and is one of the basic pillars in Data Science.  

Most of A/B tests are conducted using the statistical theory of frequentist null hypothesis statistical testing (NHST), namely t-test or z-test. \footnote{For this paper, we assume readers are already familiar with the concepts of null hypothesis statistical testing (NHST) in controlled experiments. Readers new to these concepts should refer to references such as \citet{expsurvey}.} Experimenters using NHST summarize the test result in a p-value and reject the null hypothesis $H_0$ when the p-value is less than a prescribed confidence level $\alpha$. The interpretation is that assuming all model assumptions are correct, doing so we can control the Type-I error, i.e. the probability of making a false rejection when $H_0$ is true, to be no greater than $\alpha$. 

Recently, interests in using Bayesian model comparison for two sample hypothesis testing are growing \citep{DengBayesAB, Rouder2009,Kass1995}. The type of statistical interpretations we make from Bayesian tests is fundamentally different from NHST. Under the Bayesian framework, we assume there is a prior probability $P(H_1)$ for $H_1$ (the alternative) to be true, and similarly $P(H_0)$ for $H_0$ to be true. The ratio between the two is called the prior odds. After collecting data from an experiment, we update prior odds using the Bayes Rule:
\begin{align}\label{bayesrule}
\frac{P(H_1|Data)}{P(H_0|Data)} = \frac{P(H_1)}{P(H_0)} \times \frac{P(Data|H_1)}{P(Data|H_0)}, 
\end{align}
which is commonly referred as
\begin{align*}
\text{Posterior Odds} = \text{Prior Odds} \times \text{Bayes Factor}.
\end{align*}
Note that the Bayes Factor(BF) is the likelihood ratio of observing the data between $H_1$ and $H_0$. From the posterior odds, it is straightforward to calculate the posterior probabilities $P(H_1|Data)$ and $P(H_0|Data)$. The test result can be interpreted as follows. Conditioning on a posterior odds $K$, rejecting $H_1$ will expose us to a risk of a false rejection/discovery with probability $P(H_0|Data) = 1/(K+1)$.  

In this paper we are interested in a common practice called continuous monitoring or optional stopping. This practice is best described as the following Example. 
\begin{ex}[Optional Stopping]\label{ex:cm} \ \newline
We observe data sequentially through time and at any time we can conduct statistical analysis on the data already observed. Let $t = 1, \dots, N$ be all the interim check points that we can take a peek at our A/B test results\footnote{A test result could be something like test statistics, p-value or Bayes Factor/posterior. We use this vague notion when the detail is not important.}. For any given metric M, let $R_t$ be its test result at check-point $t$. We define an event $S_t$ that is observed at time $t$ and stop the experiment at the first t such that the event happens ($R_t \in S_t$) and return result $R_t$, e.g. when we deem the result is ``significant'' or ``conclusive''. Typically event $S_t$ is defined as p-value < $\alpha$ or $P(H_0|Data) < r$. If this event didn't happen at $t=N$ we return test result $R_N$. In fact we don't even need to collect data after that. A general version could have infinite horizon.
\end{ex}

Pitfalls of continuous monitoring under NHST framework have been documented in various publications. We say the interpretation of the result is unbiased with continuous monitoring if the validity of the interpretation holds regardless of whether continuous monitoring is used. NHST is valid for fixed horizon test. But it is known to underestimate Type-I error when continuous monitoring is used. To quickly see why, if experimenters are allowed to stop the first time p-value is less than $5\%$, we will only reject more often, but no less comparing to a fixed horizon design, because the event of rejection in a fixed horizon design, i.e. only reject at time $N$, is strictly a subset of the event of rejection in the continuously monitoring design. As a result, if the Type-I error in the fixed horizon design is 5\%, the Type-I error with continuous monitoring will in general exceed 5\%. An application of the law of iterated logarithm shows when incoming data are i.i.d. continuous monitoring will inflate Type-I error to 100\% when the horizon $N$ goes to infinity, see \citet{siegmund2013sequential} and Section~\ref{sec:compare}. \citet{johari2015always} provided simulation results showing that the inflation of Type-I error is significant and could be typically above 50\% or more.   

The main result of this paper is the following theorem. 
\begin{thm}\label{mainthm} Let $\bX_t$ be all the observed data up to time $t$ and $BF_t$ be the Bayes Factor defined as $\frac{P(\bX_t|H_1)}{P(\bX_t|H_0)}$ and posterior odds $PostOdds_t$ defined as in \eqref{bayesrule} with known prior odds $P(H_1)/P(H_0)$. Let $\tau$ be any stopping time defined by a proper stopping rule, i.e. a mechanism for deciding whether to continue or stop on the basis of \emph{only the present and past events} and $\tau$ is finite almost surely. Then the interpretation of the Bayes Factor and posterior odds remains unbiased with optional stopping at $\tau$. Specifically, we have
\begin{align}\label{eq:main}
    \frac{P(H_1|PostOdds_\tau)}{P(H_0|PostOdds_\tau)} = PostOdds_\tau. 
\end{align}
\end{thm}

Theorem~\ref{mainthm} says that even though posterior odds is calculate at a random stopping time $\tau$, conditioning on observing a posterior odds of $K$, rejecting $H_1$ will expose us to a risk of a false rejection/discovery probability of $1/(K+1)$. This is a nontrivial result, because $BF_\tau$ and $PostOdds_\tau$ are calculated as if $\tau$ is a fixed time $t$. Theorem~\ref{mainthm} guarantees that the Bayesian test result remains the same interpretation even with continuous monitoring, provided that the Bayes Factor (and hence the posterior odds) are calculated using all available observations up to the stopping time $\tau$, and the stopping rule is properly defined to be based on only the present and past events. In particular, the theorem does not hold if Bayes Factor is calculated on a selected subset of the observations available at time $t$, or if the stopping rule peeked ahead into the future. These requirements are met in all common practices of continuous monitoring as in Example~\ref{ex:cm} where the stopping time is called a ``hitting time''. In conclusion, Theorem~\ref{mainthm} formally endorsed the practice of continuous monitoring in the framework of Bayesian Hypothesis Testing. This is in stark contrast to NHST, where special adjustment has to be done. Still, practices like ``re-analyze the same data using continuous monitoring after failed to reject using all data'' is not supported by Theorem~\ref{mainthm}. More bad practices are discussed later in Section~\ref{sec:bad}.

At the time of writing, there is still a lack of general agreement on whether continuous monitoring is a proper practice when Bayes test is used, more details in Section~\ref{sec:background}. The purpose of this paper is to provide arguments assessable by practitioners and engineers, while at the same time provide rigorous proofs for researchers in A/B testing community as well as related fields. With this main goal, the contributions of this paper are
\begin{compactenum}
\item We formally prove Theorem~\ref{mainthm} in Section~\ref{sec:cmproof}. 
\item We also adopt a simulation based approach to help understand Bayes Factor and posterior odds in a more tangible way, which we believe is more effective for both researchers and practitioners to understand the result and gain intuitions.
\item We discuss practical implications of using Bayesian Hypothesis Testing by comparing it to NHST. We also review and compare to NHST adjustment of continuous monitoring as in \citet{johari2015always}.
\item For practitioners, we make recommendations on when and when not to use continuous monitoring. We emphasize cases Theorem~\ref{mainthm} does not apply.
\end{compactenum}

All model assumptions required in our models are taken as granted. Although both NHST and Bayesian tests make extra model assumptions, the latter requires more such as prior and distribution under $H_1$. In practice many people use subjective priors or so called non-informative priors. These practices have been criticized a lot since there are no agreement among researchers and practitioners on which prior is appropriate. However, with the existence of rich historical A/B tests data, \citet{DengBayesAB} showed that we can learn prior objectively from the empirical data. The only assumption we are making when using historical tests data is that we assume the prior behind those historical A/B tests are the same as the current A/B test. See \citet{johnstone2004needles} for a similar technique applied in signal processing. We will provide the algorithm in Appendix~\ref{apx:prior}. But the general discussion of objective prior learning is a purely orthogonal topic to this paper and therefore out of our scope.

The rest of the paper is structured as follows. We review related work in the next Section. Section~\ref{sec:cmsim} illustrates Theorem~\ref{mainthm} using simulation studies. Proof of the main theorem is in Section~\ref{sec:cmproof}, with intuitive explanations. We emphasize bad practices where Theorem~\ref{mainthm} does not apply in Section~\ref{sec:bad}. Section~\ref{sec:compare} discuss practical implications of using Bayesian Test vs. NHST. We also review and compare to ``always valid inference'' in \citet{johari2015always}. Section~\ref{sec:con} concludes the paper with practical recommendations. 

\section{Related Work}\label{sec:background}
The need of a different theory to allow continuous monitoring in NHST framework has long been known as the subject of sequential hypothesis testing, which dates back to 1945 \citep{wald1945sequential}. Sequential hypothesis testing and later on group sequential testing have been widely used in Clinical Trials, see \citep{bartroff2012sequential} for a recent survey of the area. The idea of sequential test is only been recently popularized by \citet{johari2015always} in A/B testing community, by including it as part of the offering of the commercial A/B testing platform Optimizely. Despite it being newly introduced to A/B testing community, the theories behind sequential tests under NHST frameworks are well known by statisticians and practitioners in related areas such as clinical trials, psychology, econometrics and other social sciences. There is little dispute about the validity of the methodology. 

Bayesian hypothesis testing, on the other hand, is much less accepted and established than its frequentist counterpart. This was largely due to the need of prior knowledge that commonly requires a subjective choice or so called ``non-informative'' priors which also lack of justification. However, as argued in \citet{DengBayesAB}, for A/B testing in the big data era, much of the issue in choosing prior can be mitigated by using historical A/B tests data to empirically learn prior, assuming we have no evidence that the current experiment will have difference chance of success than the past. Putting the issue of choosing a prior aside, many Bayesians have argued that Bayesian reasoning should be immune to stopping rule. For example, \citet{dawid1979conditional} brought up the notion of conditional independence and argued that posterior based on stopping time shouldn't alter likelihood ratios. This issue is also discussed in \citet{berger1988relevance} and later \citet{Berger2004} referred to the idea as the ``stopping rule principle'' and said ``once the data have been obtained, the reasons for stopping experimentation should have no bearing on the evidence reported about unknown model parameters.'' Although the idea is well received by leading Bayesians, there are still a lot of debates going on among researchers and practitioners (who cares more about practical applications instead of entrenched debate between Bayesian and frequentist) on whether Bayesian testing and more generally Bayesian analysis is adjustment-free when optional stopping is applied. John K. Kruschke, a professor of psychological science, author of the well received book \emph{Doing Bayesian Analysis} \citep{kruschke2010doing}, made the point that Bayesian testing can be biased under optional stopping in 2013 \citep{dba13}. Andrew Gelman, a professor of statistics and political science, is on the other side and claimed that optional stopping \emph{is} Kosher in Bayesian analysis in 2014 \citep{gelman2014}. This debate is also still heated in Psychon. Bull. Rev., a journal where Bayesian hypothesis testing is relatively well received. In a 2014 paper \citet{Rouder2014} used simulation to support the case of Bayesian test with optional stopping, and to counter criticisms from \citet{erica2014decision} and \citet{sanborn2014frequentist}, both published also in 2014. 

We believe the lack of a general agreement on this issue even in mid 2010s is a clear sign that this is still a big problem for researchers and practitioners in various areas. This is especially the case for A/B testing community because 1) data are always received in near real time in a sequential fashion, 2) the technology enables and even encourages experimenters to frequently check out the test results. We found the simulation argument made in \citet{Rouder2014} to be insightful and easier to understand by practitioners and engineers than mathematical formulas. We adopt the same approach and also formally provide rigorous proof of Theorem~\ref{mainthm} in Section~\ref{sec:cmproof}.

\section{Simulation Illustration}\label{sec:cmsim}
Instead of explaining and proving Theorem~\ref{mainthm} right away, we present some simulation results first. This section serves two purposes. First, a big part of the debates about whether Bayesian test is biased with continuous monitoring is due to wrong interpretations of the Bayesian test result itself, and using frequentist measurements such as Type-I error to evaluate a Bayesian test result. This is largely due to most researchers, especially statisticians, are trained with frequentist statistics and methodologies. A correct interpretation of Bayesian posterior odds and Bayes Factor is a prerequisite for readers to understand and appreciate Theorem~\ref{mainthm} and rest of the paper. To this end, for most data scientists and engineers, we found replicable simulation results is more tangible and concrete than probability formulas. Secondly, through the simulation results, we hope readers will glimpse some intuitions on why Theorem~\ref{mainthm} is indeed expected. Readers who are not interested in rigorous proof can even skip the formal proof in the next section.

\textbf{Important:} Recall 
\begin{align*}
   \text{Posterior Odds} = \text{Prior Odds} \times \text{Bayes Factor}. 
\end{align*} Prior odds is considered known and is independent of the observations collected for the test. Prior odds is easy to interpret and interpretation of Bayes Factor, hence the posterior odds is the essence of a Bayesian test. Without loss of generality, from now on we will assume a prior odds of 1:1, \textit{i.e.} $H_1$ and $H_0$ are equally likely based on our belief without seeing any data. In this case \emph{Posterior Odds and Bayes Factor are the same.} Readers can treat them as interchangeable in this paper. 

\subsection{Bayes Factor and Fixed Horizon Test}\label{sec:fixed}
To correctly understand the meaning of Bayes Factor, we first look at the vanilla case --- a fixed horizon test first. In a fixed horizon test experimenters prescribe the total sample size, or equivalently, length and traffic of the A/B test first and only consider the test result when we collected all the data as the final result. 

We consider a simple problem of testing a normal mean. We observe N $i.i.d.$ observations $X_i, i=1,\dots,N$ from a normal distribution $N(\mu, 1)$ with unknown mean $\mu$. $\mu = 0$ under the null hypothesis $H_0$, and $\mu = \delta $ under the alternative hypothesis $H_1$. Equivalently, sample mean $\xbar$ is the sufficient statistics and it has distribution $N(0,1/N)$ under $H_0$ and $N(\delta, 1/N)$ under $H_1$. Note that this is even simpler than a A/B test because there is only one group. For two sample A/B test we replace $\xbar$ by $\Delta = \xbar_T - \xbar_C$, e.g. difference of two sample means, and the test is essentially the same as one sample test. See Appendix~\ref{apx:twosample}.   

The Bayes Factor is 
\begin{align}\label{eq:bf1}
    \frac{P(\xbar|H_1)}{P(\xbar|H_0)} = \frac{\exp(-(\xbar - \delta)^2/(2/N))}{\exp(-(\xbar)^2/(2/N))} = \exp\left(\frac{N}{2}\delta(2\xbar - \delta)\right)
\end{align}

Conditioning on observing a $\xbar$, if we plug it into Equation~\ref{eq:bf1} and get a number $K$, what does it mean? To illustrate this, we simulate $100,000$ runs and each run we simulate $N=100$ observations $X_i,i=1,\dots,N$. Since we assume prior odds 1:1, we simulate $50,000$ runs under $H_1$, where $X_i\sim N(\delta, 1)$ and the other $50,000$ runs under $H_0$ where $X_i\sim N(0,1)$. At the end of each run, we calculate Bayes Factor based on Equation~\ref{eq:bf1}. The end result of this simulation is $100,000$ Bayes Factors, half of them are from $H_1$ and half of them from $H_0$.

\begin{figure}[hbtp]
    \centering
    \includegraphics[width=0.47\textwidth]{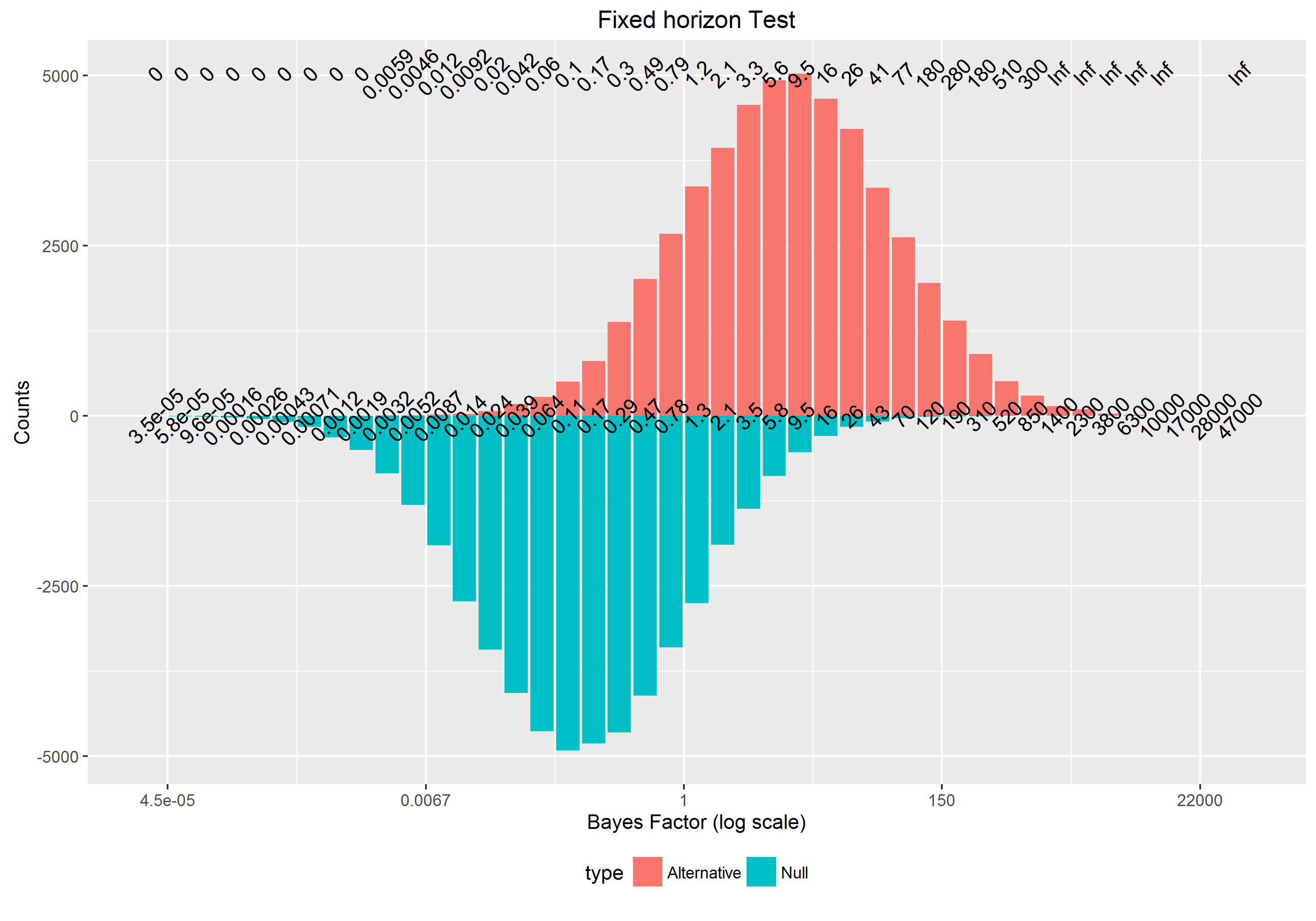}
    \caption{Histograms of Bayes Factor simulated from both models. For each bin, the number on top are the ratio of simulated Bayes Factors from alternative to those from null.}
    \label{fig:bffix}
\end{figure}

We did this simulation for $\delta = 0.2$ and $N = 100$. Figure~\ref{fig:bffix} shows histograms of those Bayes Factors in log scale, grouped by the ground truth models $H_0$ and $H_1$. Bayes Factors from $H_1$ are shown on top and those from $H_0$ are shown at the bottom. What does it mean if we observe a Bayes Factor of $2.1$? On Figure~\ref{fig:bffix}, we first group Bayes Factors close to $2.1$ together in to the same bin. There are about $4,000$ runs  from $H_1$ (height of the red bar) that produced a Bayes Factor close to $2.1$ (among $50,000$ simulation runs), while around $2,000$ (height of the blue bar) are from $H_0$ (among $50,000$ simulation runs). The actual ratio of those from $H_1$ to $H_0$ is shown on top of the plot and is $2.1$, which is the same as the Bayes Factor $2.1$ we started with. In fact, if you go through each bin carefully in Figure~\ref{fig:bffix}, you will find the number on the x-axis, which represents Bayes Factor value calculated from Equation~\ref{eq:bf1} and grouped into each bin, are very close to the actual observed count ratio of those from $H_1$ (top red) to those from $H_0$(bottom green), except those at the far tail on both sides. Is this a coincidence? Of course not. When observed a Bayes Factor of $K$, we know both model $H_0$ and $H_1$ could result in such a Bayes Factor. This simulation we did let us replay the data generation process and observe how likely it is for $H_1$ to generate such a Bayes Factor and how likely for $H_0$ respectively, which are represented, after binning similar Bayes Factors together, by the height of the top and bottom histograms. Our interest is the odds of this Bayes Factor being from $H_1$ to $H_0$, which is the ratio of heights between the red bar and the green bar. We will expect the observed ratio to be close to the true underlying odds, within some small expected error due to 1) simulation randomness and 2) discretization used in binning similar Bayes Factor together. The error from simulation randomness is smaller for those center bins, \textit{i.e.} bins where more Bayes Factors are observed from $100,000$ simulation runs, and are larger for those at the tails\footnote{Some bins on the two tails are either showing an observed ratio of 0 or Inf, for the obvious reason.}. What we observed so far can be summarized as:
\begin{align*}
    & \text{True underlying odds} = \text{Observed ratio} \\
    & = \text{Bayes Factor calculated from Equation~\ref{eq:bf1}}.
\end{align*}

This simulation illustrated two things
\begin{enumerate}
    \item Bayes Factor can be conceptually ``materialized'' as the ratio of the bar heights from the $H_1$ histogram and $H_0$ histogram. An observed Bayes Factor of $K$ means it is $K$ times more likely to be generated from $H_1$ than $H_0$. 
    \item For the fixed horizon case, Equation~\ref{eq:bf1} is the same as the true odds (at least they must be very close).
\end{enumerate}

The fact that Equation~\ref{eq:bf1} is the correct Bayes Factor is no news at all. We just followed the definition of Bayes Factor, and the observation so far confirms that we didn't make any mistake in Equation~\ref{eq:bf1}. In other words, without continuous monitoring, a fixed horizon Bayesian test of $H_1$ vs. $H_0$ as in this section using Equation~\ref{eq:bf1} provides the true underlying odds we can use to make decision. This is a special case of Theorem~\ref{mainthm} where the stopping rule is to only stop at $N$, and \eqref{eq:main} says if we calculate posterior odds which is just Bayes Factor here using $P(\bX_N|H_1)/P(\bX_N|H_0)$, it is the same as the true underlying odds conditioning on observing such a posterior odds, the left hand side of \eqref{eq:main}. We are now ready for some real stopping rules. 

\subsection{Stopping Rule Based on Bayes Factor}
If rejecting $H_0$ when observing a posterior odds no less than $K$ exposes us to a risk of false discovery at most $1/(1+K)$, a natural stopping rule is to prescribe a false discovery rate (FDR) bound and stop the test immediately if observed posterior odds already can guarantee the FDR control. We can set $K=9$ to guarantee a FDR bound of $10\% = 1/(9+1)$. 

\begin{figure}[htb]
    \centering
    \includegraphics[width=0.47\textwidth]{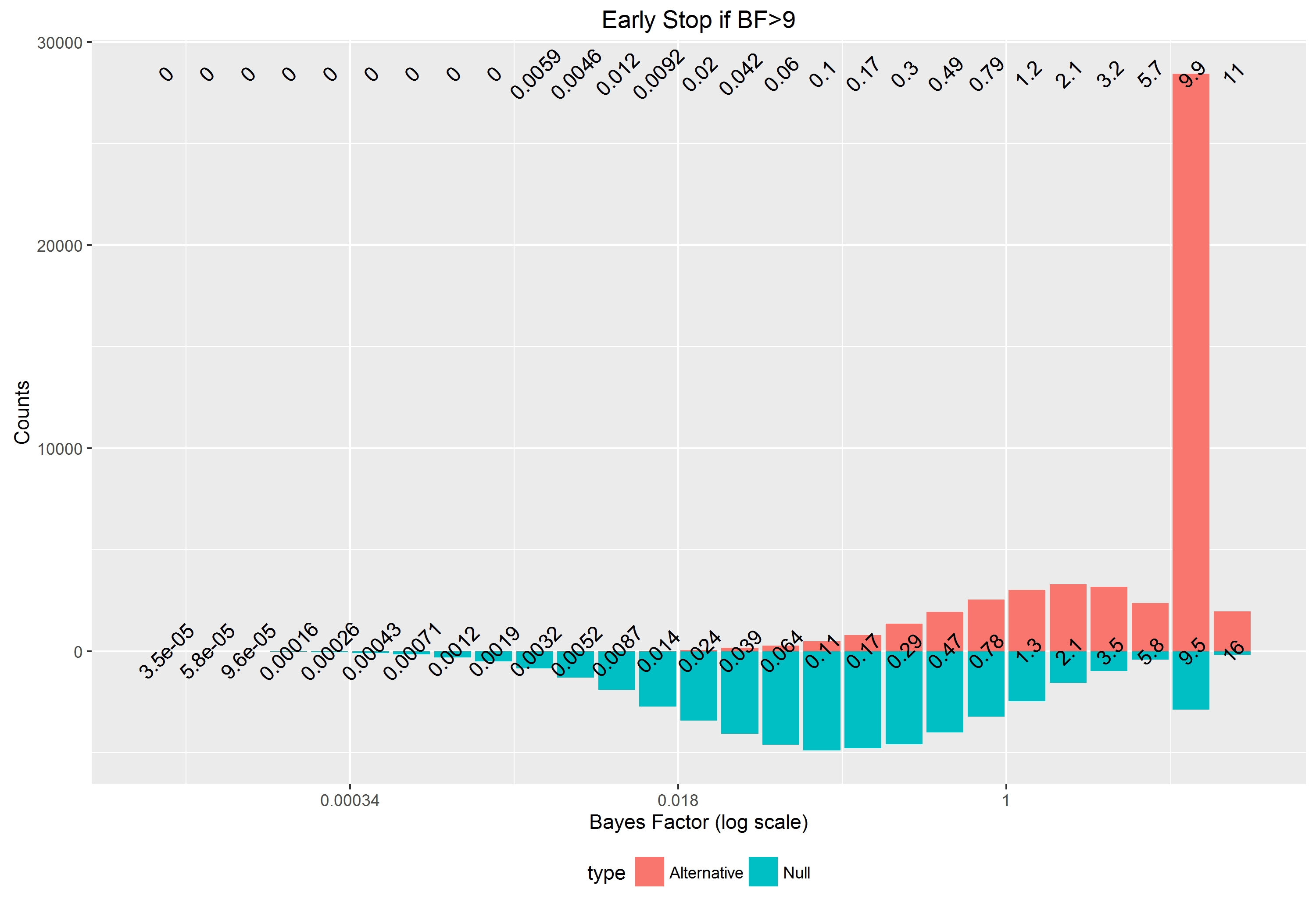}
    \includegraphics[width=0.47\textwidth]{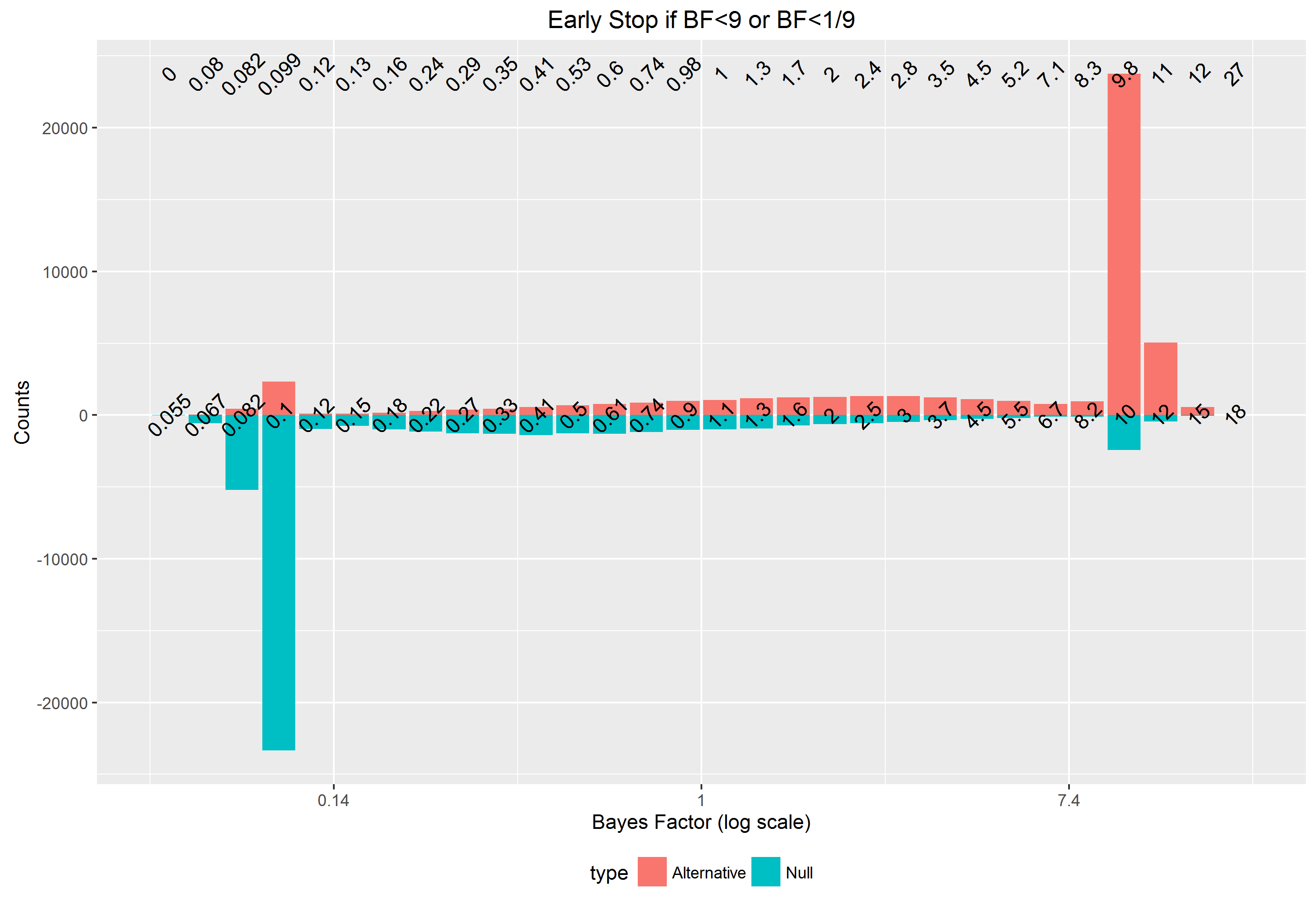}
    \caption{Histograms of simulated Bayes Factor with optional stopping. Top plot: one-sided stopping. Bottom plot: two-sided stopping.}
    \label{fig:bfstop}
\end{figure}

Similarly, we can early stop for futility and accept $H_0$ if we believe posterior of null is sufficiently large. A symmetric design is to stop if posterior odds is either no less than $K$ or no greater than $1/K$. 

Figure~\ref{fig:bfstop} illustrated both stopping rules under the same setup in fixed horizon case of Section~\ref{sec:fixed}, except for we are stopping at the first $t\le N$ when stopping rule is satisfied. In each of the $100,000$ simulation runs, regardless of whether this run is early stopped or stopped in the end, we always calculate Bayes Factor based on Equation~\ref{eq:bf1}, replacing $N$ by the observed stopping time $\tau$. Comparing Figure~\ref{fig:bfstop} to Figure~\ref{fig:bffix} shows big differences. The biggest one being the spike at the stopping Bayes Factor boundary $K=9$ (and $1/9$ for futility). However, the interesting observation is, despite the huge histogram shift for both $H_1$(red) and $H_0$(green), those numbers on the top margin --- ratios of observed Bayes Factors in each bin from $H_1$ to $H_0$, remains very close to the theoretical Bayes Factor value calculated from Equation~\ref{eq:bf1}, as in a fixed horizon test. This is exactly what Theorem~\ref{mainthm} claims, and this simulation study confirms it! 

For many who are used to the frequentist thinking of controlling Type-I error, this result seems odd. If we allow early stop, and still using the same rejection criteria of BF > 1/K, we will only reject more so we will be inflating the Type-I error. This is correct, but nonetheless does not conflict with the fact that FDR is still controlled below the designed level. 
\begin{table}[hbt]
\centering
\caption{Impact of Early Stopping}
\label{tbl:detail}
\resizebox{0.47\textwidth}{!}{%
\begin{tabular}{@{}llllll@{}}
               &        &       &       & \multicolumn{2}{l}{Early Stop Rate} \\ \midrule
               & Type-I & Power & FDR   & $H_1$            & $H_0$            \\ \midrule
Fixed Horizon & 0.018  & 0.465 & 0.037 & NA               & NA               \\
One-sided Stop & 0.060  & 0.599 & 0.09  & 59.5\%           & NA               \\
Two-sided Stop & 0.060   & 0.598 & 0.09  & 64.9\%           & 65.0\%           \\ \bottomrule
\end{tabular}
}
\end{table}

Table~\ref{tbl:detail} shows the comparison of three simulation studies we did so far in terms of Type-I error, power and FDR. In the fixed horizon design, when we reject for BF>9, the Type-I error (proportion of false rejection among the null cases) is 0.018. This value increased to 0.06 when continuous monitoring/optional stopping is introduced. Because we are rejecting more, the power of the test is also improved from 0.465 to 0.599. (Power in two-sided test is 0.598, slightly smaller than in the one-sided test. We leave it to user to figure out why.) FDR in the finite horizon cases is only 0.037, smaller than the designed bound of 0.1. This is because in the finite horizon cases a lot of rejected cases at the end of the test are actually bearing a BF much larger than the threshold 9, see Figure~\ref{fig:bffix}. This suggests that in finite horizon test, using a BF cutoff to calculate FDR might be conservative, also see \citet{efron2012large} for the differences of local FDR and FDR. When optional stopping is introduced, FDR become 0.09, very close to the designed level. The small discrepancies here is due to over-shoot, $i.e.$ we stop once BF is larger than 9 but not exactly at 9. These over-shoots are reflected in Figure~\ref{fig:bfstop} where we found a few bars beyond the spike. In large sample scenario where each individual observation won't make a big change in BF, as in most A/B tests, we can think of the time series of $BF_t$ as continuous. In this case we can stop the test with a BF almost exactly equal to 9, and the FDR will be also almost exactly 0.1. We saw that FDR control in the fixed horizon setting is conservative because we are wasting sample sizes to collect evidence beyond what we really need, and with early stopping the waste is mitigated. The last two columns in Table~\ref{tbl:detail} shows the percentage of the simulated experiment with early stopping. We saw majority of the simulated runs stopped earlier. We also calculated that the average length of the simulated runs with early stopping is about 55, much smaller than the fixed horizon of $N=100$. Based on Table~\ref{tbl:detail}, one could argue that early stopping is always superior than the fixed horizon test, and should be recommended. More discussions are in Section~\ref{sec:compare}.

\subsection{General Stopping Rule}
Theorem~\ref{mainthm} holdes for general stopping rules, not only those based on BF cutoff values. For experimenters who want to ``hack'' p-values, they could choose to stop once p-value is less than $\alpha$. Here we did the simulation study with the stopping rule with both criteria: 1) p-value less than 0.1, and 2) the sample sizes is at least 10. 

\begin{figure}[hbt]
    \centering
    \includegraphics[width=0.47\textwidth]{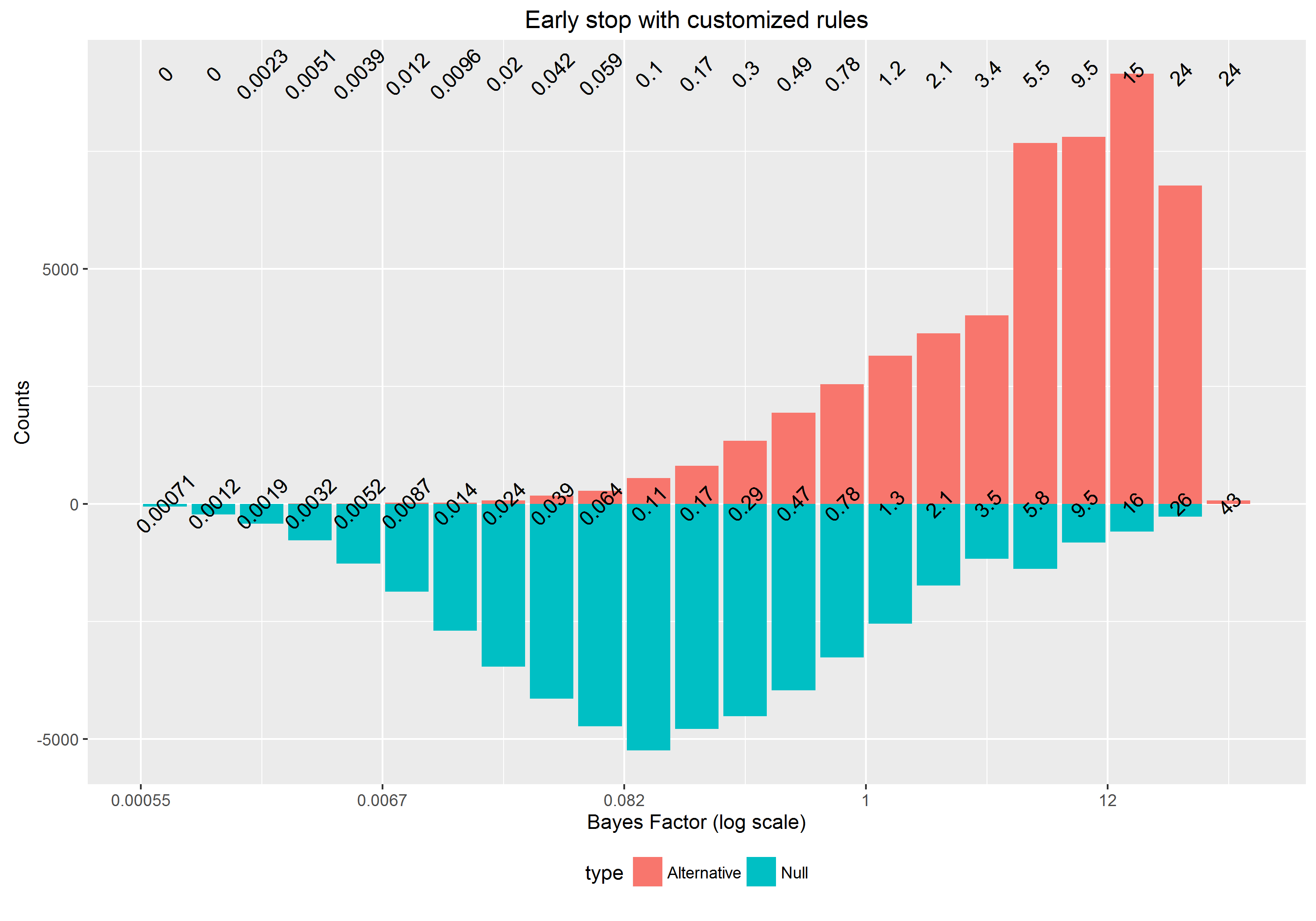}
    \caption{Using a stopping rule based on p-value and minimum required sample size.}
    \label{fig:pvaluerule}
\end{figure}

Figure~\ref{fig:pvaluerule} shows the simulation results, this time with a rather bizarre histogram for $H_1$ runs. The important part is the observed actual ratio on the top margin still closely tracks the theoretical Bayes Factor values on the x-axis. 

\subsection{Composite Alternative}\label{sec:simcomp}
So far in this section we have been using a overly simple alternative model $H_1$ where the treatment effect is assumed to be known. This is not very realistic since we never know the effect so that alternative is always a composite alternative where $\delta$ can be anything nonzero. In Bayesian model comparison we need to put a prior distribution for $\delta$ under $H_1$, in addition to the prior odds. Following \citep{johari2015always} and \citep{DengBayesAB}, we put a normal prior $N(0, \sigma_0^2))$. Under this $H_1$, $X_i \sim N(0, \sigma_0^2+1)$ and the formula for Bayes Factor assuming a fixed sample size $N$ changes to 
\begin{align}\label{eq:composite}
   \frac{N(\xbar; 0, \sigma_0^2+1/N)}{ N(\xbar;0,1/N)} 
\end{align}

\begin{figure}[htb]
    \centering
    \includegraphics[width=0.47\textwidth]{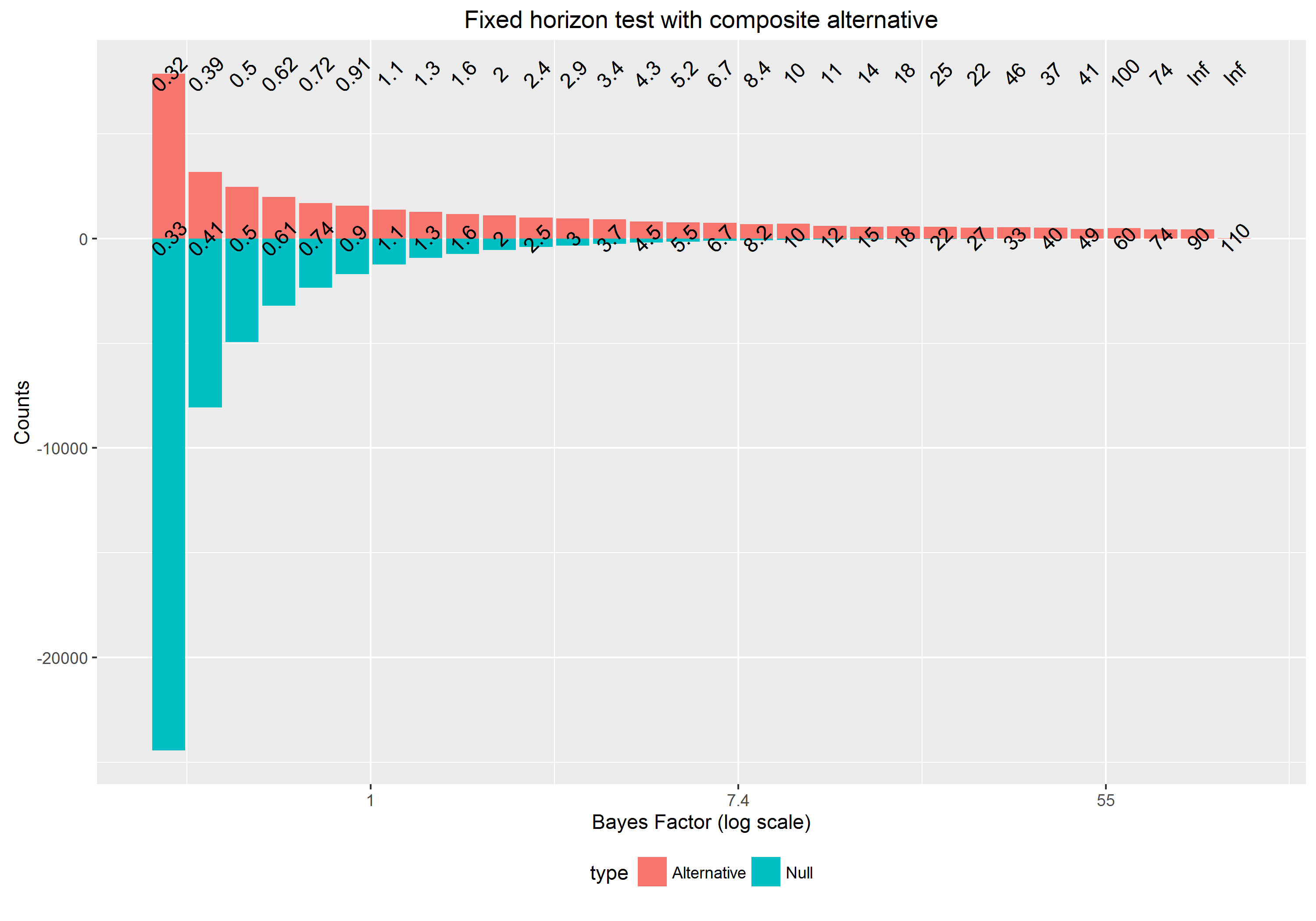}
    \includegraphics[width=0.47\textwidth]{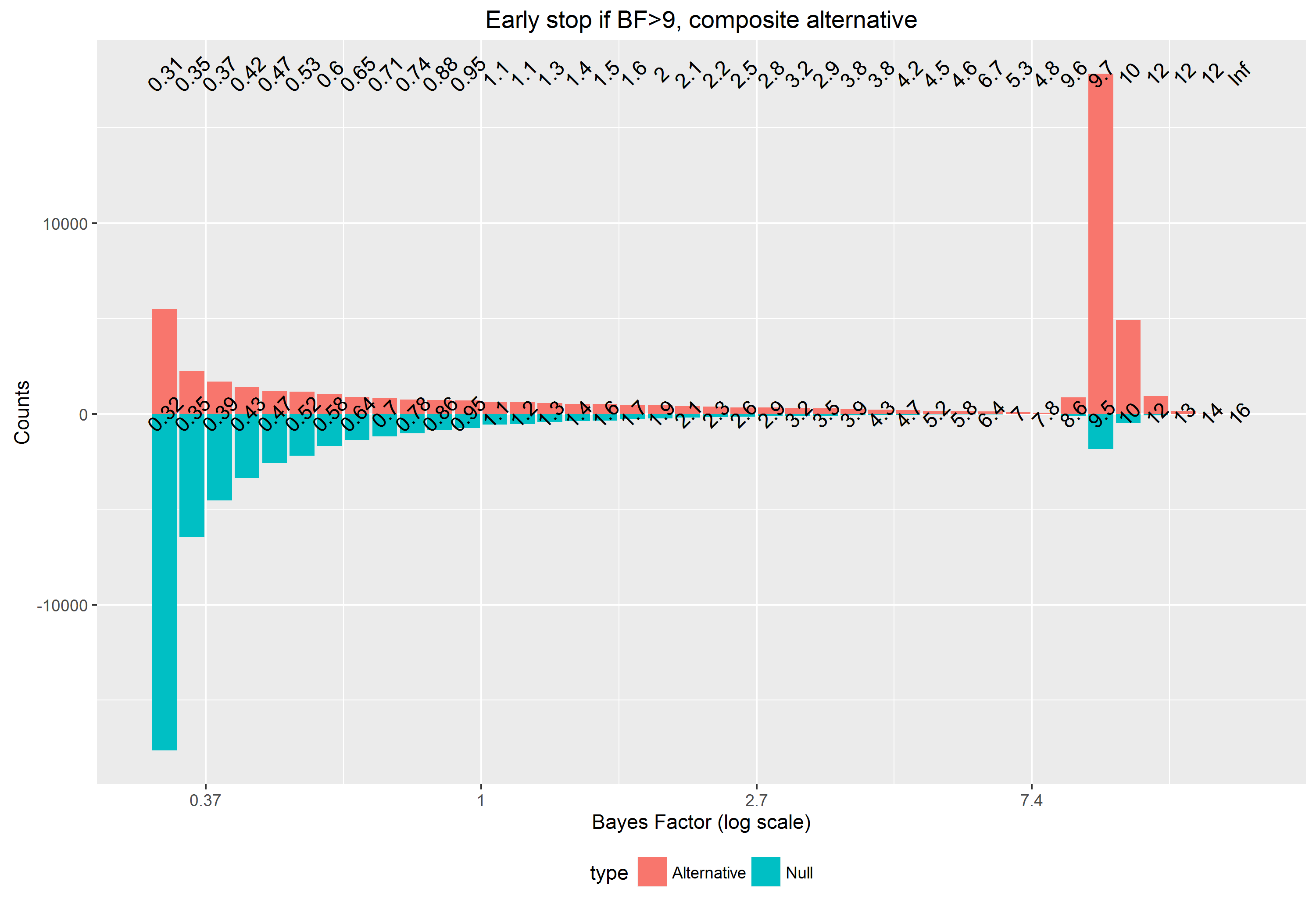}
    \caption{Composite $H_1$ with Normal prior. Top: fixed horizon(x-axis limit to 100). Bottom: early stopping when BF>9. }
    \label{fig:composite}
\end{figure}

A similar simulation to those above in this section is run by setting $N = 1,000$. We also set $\sigma_0 = 0.1$ to generate $50,000$ independent $\delta$ first for each of the simulation runs from $H_1$. At the end of each runs(or at the stopping time) we compute Bayes Factor based on \eqref{eq:composite} with $N$ for fixed horizon setting or $\tau$ in its place when optional stopping is introduced. Figure~\ref{fig:composite} shows the results for both fixed horizon setting and optional stopping with BF cutoff at 9. In the fixed horizon setting, the histogram is much more dispersed than the previous precise alternative case. Some BF is as large as several thousands so we only show those no greater than 100. Early stopping effectively eliminated those extremely large BF, creating spikes around 9. We hope readers at this point already noticed that the top margin numbers are very close to the theoretical BF values on the x-axis.

\section{Proof of Main Theorem}\label{sec:cmproof} 
We now prove Theorem~\ref{mainthm}. Readers who only need intuition are recommended to skip the proof and jump to Section~\ref{sec:intuition}. Both sides of the \eqref{eq:main} are random variables depending on $PostOdds_\tau$. It is equivalent to the following: 
\begin{align*}
   & \frac{P(H_1|PostOdds_\tau = K)}{P(H_0|PostOdds_\tau = K)} = \frac{P(H_1 \text{ and } PostOdds_\tau = K)}{P(H_0\text{ and } PostOdds_\tau = K)}\\
   & =   \frac{P(PostOdds_\tau = K|H_1)}{P(PostOdds_\tau = K|H_0)}\times \frac{P(H_1)}{P(H_0)} = K, 
\end{align*}
for any K where $P(PostOdds_\tau = K)>0$. Let $K'= K\times P(H_0)/P(H_1)$. The event $\{PostOdds_\tau = K\}$ is equivalent to $\{ BF_\tau = K' \}$. The last equality above after rearranging the prior odds $P(H_1)/P(H_0)$ to the right side becomes
\begin{align}\label{eq:bfeq}
    \frac{P(BF_\tau = K'|H_1)}{P(BF_\tau = K'|H_0)} = K'.
\end{align}
Without loss of generality, we only need to prove \eqref{eq:bfeq}. 

We first prove for the fixed horizon case, which is a direct result of \emph{likelihood ratio identity}, or \emph{change of measure}. For any fixed $t$, let $\q_t = P(\cdot|H_1)$ and $\p_t = P(\cdot|H_0)$ be the probability measure under $H_1$ and $H_0$ respectively for observations up to $t$, both have a density function w.r.t. to Lebesgue measure on the real line. Let $A$ be any event observable at time $t$.\footnote{$A\in \fcal_t$ where $\fcal_t$ represents the set of measurable events at time $t$. $\fcal_t$ is called a filtration because $\fcal_s \subseteq \fcal_t$ for any $s\le t$.} The likelihood ratio identity\footnote{It is also called change of measure identity because the equation transforms an expectation under a measure $\p$ into an expectation under another measure $\q$. This is a special case of the Radon-Nykodym Theorem in measure theory.} ensures
\begin{align*}
   \q_t(A) = E_{\p_t}\left(\indicator_{A} \times \frac{d\q_t}{d\p_t}\right),
\end{align*}
where $d\q_t/d\p_t$ is the likelihood ratio and $\indicator_{A}$ is the binary indicator function for event $A$. Recall Bayes Factor is defined as the likelihood ratio. Replace $d\q_t/d\p_t$ by $BF_t$, set $A = \{ BF_t = K'\}$ in the above to get
\begin{align}
    \q_t(BF_t = K') = E_{\p_t}(\indicator_{A})\times K' = K' \times \p_t(BF_t = K'), 
\end{align}
which is \eqref{eq:bfeq}. 

We can generalize this argument for random time $\tau$. Theorem~\ref{mainthm} requires $\tau$ to be a stopping time\footnote{$\tau$ is a stopping time w.r.t. to a filtration $\fil_t$ if $\{\tau \le t \} \in \fcal_t$.} so that the event $\{ \tau = t\}$ is observable at time $t$. This is a necessary requirement to ensure that we can apply the likelihood ratio identity for the event $\{ BF_t = K' \text{ and } \tau = t\}$(observable at $t$) to get
\begin{align}\label{eq:bfeq2}
    \q_t(BF_t = K',\tau = t) = K' \times \p_t(BF_t = K',\tau = t).
\end{align}
If $\tau$ can only take value from $1$ to a maximum horizon $N$ (experiment stop at $N$ no matter what, which covers all practical cases), summing up \eqref{eq:bfeq2} over all $t$ entails
\begin{align*}
    &P(BF_\tau = K'|H_1) = \sum_{t=1}^N P(BF_\tau = K', \tau=t |H_1) \\
    & =\sum_{t=1}^N \q_t(BF_t = K',\tau=t) = \sum_{t=1}^N K'\p_t(BF_t = K',\tau=t)\\
    & = \sum_{t=1}^N K'P(BF_\tau = K',\tau=t|H_0) = K' P(BF_\tau = K'|H_0)
\end{align*}
which is \eqref{eq:bfeq} and the proof is completed. Notice how we changed $BF_\tau = K'$ to $BF_t = K'$ once we restrict ourselves to the set $\tau = t$ in the second and fifth equality. The essence of the proof is to show
\begin{align}\label{eq:intuition}
    \frac{P(BF_\tau = K',\tau=t|H_1)}{P(BF_\tau = K',\tau=t|H_0)} = K'
\end{align}
for every $t\le N$ by applying likelihood ratio identity \eqref{eq:bfeq2} and then sum up both numerator and denominator of \eqref{eq:intuition} over $t$ to recover \eqref{eq:bfeq}. For potentially unbounded $\tau$, we just need to sum up to infinity and the result still holds because the sum series of both numerator and denominator of \eqref{eq:intuition} are finite. 

\textbf{Important Remark}. We make the remark that Theorem~\ref{mainthm} \textbf{does not} require observation $\bX$ to be sequential \textit{i.i.d.} observations as in earlier simulation examples. All we used in the proof is the likelihood ratio identity for \emph{the whole path of observations $\bX_t$ up to $t$ for any $t$}. In a typical A/B test with user level tracking, users who visit the site multiple times will provide multiple observations sequentially. Since there is a strong between-user correlation, when we look at $\bX_t$, which includes all sequential observations (page-views) from different users, they are not independent. However, at any given time $t$, we can always first aggregate $\bX_t$ to the randomization unit, which is user. Take the metric Revenue per user as example, $\bX_t$ is the sequence of revenues for each page-view up to time $t$. For each user, we can sum up revenues as $Y_{it}, i=1,\dots,N_t$ where $N_t$ is number of unique users. We can treat $Y_{it}$ as \textit{i.i.d.} when computing likelihoods under both $H_1$ and $H_0$. For ratio metrics such as Click-Through-Rate(CTR), $Y_{it}$ can either be CTR for each user and average of $Y_{it}$ is the average CTR over all users --- a double average metric, or $Y_{it}$ can be a pair $(Clicks_{it},PageViews_{it})$ and the metric is the sum of clicks over all users divided by sum of page-views. Delta Method is required to compute the likelihood for the latter case. The main point is that the original sequential observations might not be \textit{i.i.d.} but after aggregated to randomization unit level, likelihood ratio can be easily calculated using those aggregated values which can be assumed \textit{i.i.d.} because of the randomization design.

\subsection{Intuitive Explanation}\label{sec:intuition}
There is an intuitive explanation of Theorem~\ref{mainthm} using the same simulation procedure in Section~\ref{sec:cmsim} as a thought experiment. Again we assume prior odds is 1:1 so Posterior Odds equals to the Bayes Factor and \eqref{eq:main} becomes \eqref{eq:bfeq}. 

We simulate $M$ paths from $H_1$ and $H_0$. $M$ is a very large number, almost infinite. So every path simulated from $H_1$ which has nonzero probability under $H_0$ will have the same path simulated under $H_0$, and vice versa. For each path we simulate the whole path $\bX$ up to the fixed final horizon $N$. For any $t\le N$ and any path $\bX$, we can calculate a Bayes Factor at time $t$ to be $BF(\bX_t)$ as likelihood ratio $P(\bX_t|H_1)/P(\bX_t|H_0)$. No stopping rule has been introduced yet so everything so far belongs to the fixed horizon case. Define $Path(\bX_t|H_i), i=0,1$ to be the set of all paths simulated from $H_i,i=0,1$ having the same subpath $\bX_t$ up to time $t$, and let $|Path(\cdot)|$ denote the total number of paths in a path set, \textit{i.e.} cardinality. Then \emph{for any $t$ and any path $\bX$ with subpath $\bX_t$, $|Path(\bX_t|H_1)|/|Path(\bX_t|H_0)| = BF(\bX_t)$.} Intuitively, this means for every subpath $\bX_t$ simulated from $H_0$, there are on average $BF(\bX_t)$ \emph{exaxt} subpaths simulated from $H_1$. Because this statement is true for any subpath. If we only look at subpaths such that $BF(\bX_t) = K$, we have $|Path(\{\bX_t:BF(\bX_t)=K \}|H_1)|/|Path(\{\bX_t:BF(\bX_t)=K \}|H_0)|= K$, for any $K$. 

Now we introduce stopping rule. Pick any path $\bX$ simulated from $H_0$, say the stopping rule will stop at time $t$ and we computed the Bayes Factor to be $K$. The previous argument shows there will be on average $K$ number of the the same exact subpath simulated from $H_1$. Here comes the important part! \emph{Because the stopping rule does not depend on observations after the stopping time, all subpaths simulated having the exact same subpath $\bX_t$ up to $t$ will also have the exact same stopping time at $t$!}(See the next section for examples of bad stopping rules where this property is not true, hence Theorem~\ref{mainthm} does not apply.) After we gathered all paths simulated from $H_0$ with the same Bayes Factor $K$ at time $t$ which also stopped at time $t$ according to the stopping rule, for each one of them we can find $K$ exact same subpaths which \emph{also stopped at time $t$}. By one more step of gathering all such set of paths for every possible $t\le N$, it is then intuitively clear that the number of paths gathered together from $H_1$ and $H_0$ have a ratio of exactly $K$. This is exactly what we tried to demonstrate via various simulations in Section~\ref{sec:cmsim}. 

\section{Bad Practices}\label{sec:bad}
Theorem~\ref{mainthm} is a general result with very mild assumptions which are satisfied in most cases. But failure of satisfying those assumptions can result invalid test results. We list three bad practices so readers can be aware of the limitations of the result in this paper. One critical assumption is that the stopping rule is properly defined that only uses information already observed, without peeking into the future. One example for an improper stopping rule is to reassess all the observations at some time $t'$, and then decide to only use the data up to an earlier time $t<t'$, e.g. stop at $t$ after seeing data at a later time $t'$. This practice is called \emph{data snooping} and is not supported by Theorem~\ref{mainthm}. There are two common bad practices related to data snooping:
\begin{ex}[Re-analysis after Fail to Reject]\ \newline
Finite horizon test at $N$ failed to reject $H_0$. The same data is then reanalyzed using continuous monitoring as in Example~\ref{ex:cm}.
\end{ex}
\begin{ex}[Optimal Stopping]\label{ex:optimalstop} \ \newline
The basic setup is the same as in Example~\ref{ex:cm}. This time we first collect all the data up to finite horizon $N$. Then, we look at our data and try to find the \emph{best} check-point $t$ so the test result $R_t$ is the most favorable. The difference between this example and continuous monitoring is that for the latter the decision of stopping the experiment is made without peeking at the data in \emph{future}. 
\end{ex}
In both examples above, if we collected all the data up to horizon $N$ and \emph{did the test}, we should always report this test result instead of re-analysing the data or try to cherry pick the optimal stopping time. This is because Bayesian test is consistent: as we observe more data, posterior $P(H_1|Data)$ converges to 1 if $H_1$ is true and to 0 otherwise. This means we should always prefer the decision made from more data. However, it \emph{is} possible that continuous monitoring might have rejected $H_1$ (if it were used) but the finite horizon test at $N$ does not. Does that mean continuous monitoring makes more error? No! Table~\ref{tbl:detail} shows continuous monitoring does increase the amount of null rejection, without sacrificing the false discovery rate. In the above case the posterior odds realized in the end of the experiment at horizon $N$ shows $H_1$ is unlikely to be true. It is fair to say \emph{if we had been using continuous monitoring, we would likely be making a false discovery at that time, based on newer observations}. However, let $K'$ be the posterior odds reported by continuous monitoring, Theorem~\ref{mainthm} guarantees that it is $K'$ to 1 odds that we will see posterior odds increasing to $\infty$ if we keep getting more data, than decreasing to $0$. In other words, it is $K'-1$ more likely our decision still uphold in the end of the experiment, than reversed as in the hypothetical case. 

Another critical assumption in Theorem~\ref{mainthm} is that the likelihood ratio has to be correctly calculated with all available observations, \textit{i.e.} the whole subpath $\bX_t$. In particular, we cannot cherry pick only those observations that favors one hypothesis. Here is another bad example which happens a lot in practice. 
\begin{ex}[Continuous Testing until Win]\label{ex:conttest} \ \newline
With agile development and continuous A/B testing, a team can iteratively modifying and testing a feature until seeing a successful test result.     
\end{ex}
In NHST, even if the feature has no effect, there is still $\alpha$(typically 0.05) chance that the result could be statistically significant. This means for every 20 iterations, we might just declare a success without really having any true effect. This is like continuous monitoring, but the difference is that here each new test only uses its own data. Using Bayesian test, if we need to calculate the likelihood ratio up to the $t$-th test, we have to aggregate all the evidence together, not just looking at the last one. If all tests are independent replications of the same test, aggregating evidence in Bayes test is trivial, we just need to multiply likelihood ratio for each of the replications all together. This way even if we might have a few large likelihood ratio favoring $H_1$, but if $H_0$ is the ground truth there have to be more smaller likelihood ratios so the product is small. In fact it will converge to $0$ if we keep doing replication runs. In practice, since iteration runs are not exactly replications, it is still a challenge how we should properly aggregate evidence from multiple experiment runs together. Ignoring the prior runs can still result in more false discovery than nominally controlled. Technically this is the area of multiple testing and selection bias. See \citet{ludeng2016} for some preliminary results. 

\section{Compare to Frequentist Methods}\label{sec:compare}
\subsection{NHST}
We compare Bayes testing to NHST in this section to reveal why continuous monitoring is an issue for NHST but not for Bayes testing. For simplicity, we assume $X_i, i=1,\dots$ are sequential \textit{i.i.d.} observations from $N(\mu, 1)$.  

First we test $H_0:\mu=0$ against $H_1:\mu=\delta$. In NHST we reject $H_0$ when the z-statistics is larger than a constant threshold, while in Bayes testing we reject when the Bayes Factor \eqref{eq:bf1} is larger than a constant threshold. Simplifying both rejection boundary resulted in
\begin{align}\label{eq:boundary}
    |\sqrt{n} \xbar_n| & > C_1 \quad \text{NHST},\\
    |\sqrt{n}\xbar_n|& > C_2\sqrt{n} + C_3/\sqrt{n} \quad \text{Bayesian},
\end{align}
where $C_i,i=1,2,3$ are constants. Asymptotically, NHST reject when test statistics is larger than $O(1)$ and the Bayesian test's rejection boundary is $O(\sqrt{n})$. Why $O(\sqrt{n})$? The reason of having a $O(\sqrt{n})$ in this case is due to the fact that we put a precise alternative $\mu = \delta$ for $H_1$. If $H_1$ is true, by central limit theorem, we would expect most of times we should observe $\xbar$ within $1/\sqrt{n}$ neighborhood of $\delta$. If $H_0$ is true, then $\xbar$ should be within $1/\sqrt{n}$ neighborhood of $0$. In NHST, we only assess the likelihood of $H_0$, so we reject if $\xbar$ is outside of $1/\sqrt{n}$ neighborhood of $0$. Bayesian testing is symmetric, \textit{i.e.}, we compare the likelihood of $H_1$ vs $H_0$. In this case since both hypothesis expect observation to be within their $1/\sqrt{n}$ neighborhood, the ``classification'' line should naturally be for $\xbar$ at the midway $\delta/2$, which means $\delta/2\times\sqrt{n}$ for z-statistics. 

Is a rejection boundary of $O(\sqrt{n})$ means Bayesian test will be much less sensitive than NHST? Not necessarily. First of all, if $H_1$ is indeed true and $X_i$ will have mean $\delta$, then the chance that we observe $\xbar$ within $\sqrt{n}$ neighborhood of $\delta$ is high, meaning it should not be too much problem for the z-statistics to breach the $O(\sqrt{n})$ boundary. Secondly, if we replace the precise alternative to a composite alternative with a normal prior $N(0,V^2)$ on $\mu$ under $H_1$, similar calculation reveals an asymptotic rejection boundary of $O(\sqrt{\log(n)})$, very close and almost can be considered $O(1)$ in practice\footnote{Sample size in practice are typically restricted to 1\% of all traffic to 100\%, only an order of magnitude of 2.}.  

But why not simply $O(1)$? Law of iterative logarithm(LIL) shows in the \textit{i.i.d.} sequential $X_i$ case $\sqrt{n}|\xbar|$ even under $H_0$ will breach the boundary $C\sqrt{\log(\log(n))}$ \emph{infinitely often for any $C$}. Any rejection boundary that is not at least $O\left(\sqrt{\log(\log(n))}\right)$ will be breached with probability $1$ under $H_0$, and with Type-I error $1$ with continuous monitoring, when $n\to \infty$. If Bayesian test have a rejection boundary not bigger than $O\left(\sqrt{\log(\log(n))}\right)$, it will reject $H_0$ every time, and the false discovery rate is solely determined by the prior odds. This is in contradiction to the result of this paper. In other words, because of Theorem~\ref{mainthm}, any valid Bayes test will have a rejection boundary larger than $O\left(\sqrt{\log(\log(n))}\right)$. 

In practice, it is true that Bayesian tests are in general more conservative than NHST. Partly it is because the rejection boundary need to be at least $O\left(\sqrt{\log(\log(n))}\right)$---a price to pay for continuous monitoring. Another more important part is because Prior Odds $P(H_1)/P(H_0)$ usually favors $H_0$, and sometimes a lot, \textit{e.g.} knowing certain metric rarely truly moved. 

\subsection{Sequential Test}
Since fixed horizon NHST does not control Type-I error well, corrections have to be made under frequentist framework to keep controlling Type-I error at desired level. \citet{wald1945sequential} first introduced sequential probability ratio test (SPRT). When assuming both $H_0$ and $H_1$ are precise, e.g. simple alternative, SPRT make decision based on the likelihood ratio $P(Data|H_1)/P(Data|H_0)$ and reject $H_0$ when $LR>B$ and accept $H_0$ when $LR<A$. The bounds $A$ and $B$ can be chosen such that Type-I error and Type-II error can be controlled at desired levels. In practice we usually don't know the exact treatment effect under the alternative $H_1$. \citet{lai1988nearly} extended the idea of generalized likelihood ratio test into sequential setting. \citet{johari2015always} introduced the notion of ``always valid inference'', where they used a variant of Wald's SPRT with a normal distribution on $H_1$. It is called mSPRT (m stands for \emph{mixture}). From a Bayesian test perspective, this is equivalent to putting a normal prior for the average effect size $\mu$ under $H_1$, same as in Section~\ref{sec:simcomp}, also see Appendix~\ref{apx:prior}. However, despite the similarity in the form of likelihood ratio there are two main differences:
\begin{compactenum}
    \item mSPRT does not take prior knowledge into account, while Bayesian test encode these information as prior odds $P(H_1)/P(H_0)$. We know from experience, and from historical that some metrics are easier to move than the others. For mSPRT the rejection boundary for all metrics are the same. In A/B testing, it is reported that most ideas fail to deliver desired movement, or even move the success metric. Prior odds $P(H_1)/P(H_0)$ for most metrics are less than 20\%. 
    \item A more fundamental difference is the interpretation of the results. mSPRT controls Type-I error --- the chance of false rejection when $H_0$ is true, while Bayesian test controls False Discovery rate (FDR) --- the chance of false rejection when decided to reject $H_0$. There is no simple relation ship between the two. When we reject more aggressively, Type-I error will increase, but FDR does not necessarily increase, as long as more aggressive rejection will also reject more true positives. FDR also extends to multiple testing cases easier than Type-I error\footnote{FDR was originally brought up in frequentist multiple testing scenario. FDR does not make much sense in frequentist framework for single test because $H_0$ for a test is assumed to be either true or false.  FDR is 1 if $H_0$ is true and 0 otherwise. There is no room in the middle with prior probability.}. \citet{johari2015always} also proposed a way to generalize the classic Benjamini-Hochberg procedure \citep{benjamini1995controlling} to sequential setting to control FDR of mSPRT tests in multiple testing scenario. This additional step is not necessary in Bayesian test. 
\end{compactenum}
For both mSPRT and Bayesian test with normal alternative model under $H_1$, the asymptotic rejection boundary for test statistics are both $O\left(\sqrt{\log(n)}\right)$. Both methods makes the same assumption about the normal alternative model. Bayesian test makes extra assumption that prior odds are also known. To use either method, we strongly suggest learning parameters from empirical data, as explained in appendix, see \citet{DengBayesAB} for more detail.

\section{Conclusion \& Recommendation}\label{sec:con}
With Theorem~\ref{mainthm} rigorously proved, we believe the debate over whether continuous monitoring is a valid practice when Bayesian Hypothesis Testing is used should be settled. The answer is unequivocally yes, in the sense that the Bayesian posterior remains \emph{unbiased} \eqref{eq:main} when a \emph{proper} stopping rule is used. We emphasize that the correct understanding of \eqref{eq:main} and interpretation of Bayesian test result as controlling FDR is critical. Many researchers, even some versed in Bayesian statistics \citep{kruschke2010doing}, made the mistake of evaluating Bayesian test results conditioning on either null is true or alternative is true. The correct Bayesian interpretation always requires a prior odds weighing the alternative and the null. Our simulation illustrations in Section~\ref{sec:cmsim} serve the very goal of helping readers understand this crucial point in a non-technical way. 

Two natural questions are then raised by practitioners. 1) Because the fundamental difference in the statistical conclusions we can make from NHST and Bayesian test, which one shall we use in practice? 2) Is the result of this paper telling us when we use Bayesian test we should always use continuous monitoring? 

For the first question, we focus on the difference of controlling Type-I error and FDR. If false rejection of any single test cost us a lot, and the cost of false rejection is considered higher than false negative(fail to reject true $H_1$), then Type-I error seems to be a better criterion to control. A good example for this case is clinical trial. If our goal is not focus on each individual test, but the overall performance of our decision on a large set of tests, and the cost of false rejection and false negative are in the same order, then we believe FDR is a better criterion. Large scale A/B testing platform is an example of the latter \citep{abScale}. In an agile environment, where success are built upon a lot of small gains, as long as we are shipping more good features really meet customer needs than useless ones, we are moving in the right direction. 

For the second question, continuous monitoring is not always recommended. In many cases, the goal of the experiment is not only to confirm the existence of the treatment effect, but also to measure it. In A/B tests, it is not uncommon for a feature to have time-varying treatment effect such as weekday and weekend differences. To capture the weekly cycle, running tests for a whole week or multiple of weeks are often recommended. It is also possible that the treatment effect only exists in the weekend and we might early stop the experiment during the first few weekdays when we use continuous monitoring with an early acceptance of $H_0$. But continuous monitoring should be recommended in many other scenarios. Shutdown a bad experiment is one application. We want to shutdown an experiment once we have enough evidence that the treatment is giving user a very bad experience. Another example is comparing a few closely related alternative candidates, e.g. tuning parameters for a backend algorithm, in which case we might assume the \emph{ordering} of the treatment effects won't be time-varying and hence we can early stop inferior candidates and ramping up outperforming candidates based on Bayesian posterior. The last example is studied in more detail in the literature of multi-armed bandit and Thompson sampling \citep{chapelle2011empirical} and the result of this paper justifies Thompson sampling for using Bayesian posterior to dynamically change data gathering.   


{\small
\setlength{\bibsep}{0.0pt}
\bibliography{library}
}

\appendix
\section{Two Sample Test as One Sample Test}\label{apx:twosample}
We review two sample t-test and its large sample z-test version. We also show how to ``normalize'' the scale and turn the two sample test into an equivalent one sample test. This transformation also introduce the notions of \textbf{effect size} and \textbf{effective sample size} which is important in Bayesian test. 

Suppose observations for treatment and control groups are i.i.d. from two distributions with unknown mean $\tau_T$ and $\tau_C$ respectively. Denote our observations by $Y_i,i=1,\dots, N_T$ and $X_i, i=1,\dots, N_C$. We test the null hypothesis $H_0: \tau_T-\tau_C = 0$ against the alternative $H_1: \tau_T\neq \tau_C$. Without assuming distributions of $X$ and $Y$, we use the central limit theorem and hence use Wald test which is large sample version of the well-known t-test. The test statistic is 
\begin{align*}
\text{Z} := \frac{\xbar-\ybar}{\sqrt{\sigma_T^2/N_T+\sigma_C^2/N_C}} = \frac{\Delta}{\sqrt{\sigma_T^2/N_T+\sigma_C^2/N_C}},
\end{align*}
where $\sigma_C$ and $\sigma_T$ are variances of $X$ and $Y$. The variances are also unknown but in large sample scenario we assume they are known and use their estimates. Note that metrics are often in different scales. We first define $N_{E} = 1/(1/N_T+1/N_C)$ to be the \textbf{effective sample size}. And then let $\sigma^2$ be the \textbf{pooled variance} such that $\sigma^2/N_{E} = \sigma_T^2/N_T+\sigma_C^2/N_C$. With $\delta = \Delta/\sigma$, Z-statistics can be rewritten as
\begin{align}\label{eq:zformula}
\text{Z} = \frac{\delta}{\sqrt{\sigma^2/N_{E}}}.
\end{align}
$\delta$ is $\Delta$ scaled by pooled standard deviation and is called the \textbf{effect size}. Finally, define
\begin{align}
\mu := E(\delta) = E(\Delta)/\sigma = (\tau_T-\tau_C)/\sigma
\end{align}
is the average treatment effect scaled by $\sigma$. When $\sigma$ is known, inference on $\tau_T-\tau_C$ and $\mu$ are equivalent. In Bayesian analysis it is common to define prior for $\mu$ as it is scaleless. 

\section{Objective Prior Learning}\label{apx:prior}
Recall $\mu$ is the average effect size. Under $H_0$, $\mu=0$. Under $H_1$, we assume a prior $\pi$ for $\mu$. For both cases we observe $\delta \sim N(\mu, 1/N_E)$. In addition, we assume a prior probability $p$ for $H_1$ being true, and also under $H_1$, $\pi \sim N(0,V^2)$ for some $V$. Our challenge is to learn both $p$ and $V$ without the need of subjectively assigning one. 

Here we take advantages of historical experiment results and use them to learn the prior. Suppose for a given metric, we have $N$ previously conducted tests with observed effect size and effective sample size $(\delta_i, N_{Ei}),i=1,\dots,N$. We have no idea which of those are from $H_0$ or $H_1$. Fitting the model to find MLE isn't straightforward, due to the fact that we don't know each $\delta_i$ belongs to $H_0$ or $H_1$. Fortunately, a solution for this type of hidden latent variable problem, called Expectation-Maximization, is well-known\citep{Dempster1977}. EM algorithm in our case reduces to a fairly intuitive form as the following. 

\noindent Step I. If $p$ and $V$ are known, the posterior odds for each $\delta_i$ belonging to $H_1$ against $H_0$ have the simple form
\begin{align}\label{eq:postodds}
\frac{\phi(\delta_i;0, 1/N_{Ei}+V^2)}{\phi(\delta_i;0,1/N_{Ei})} \times \frac{p}{1-p}
\end{align}
where $\phi(x;\mu,\sigma^2)$ is the normal density with mean $\mu$ and variance $\sigma^2$. Convert posterior odds to $P_i:=P(H_1|\delta_i; p, V)$.

\noindent Step II. Set $p$ to be $\overline{P(H_1|\delta; p, V)}$ by taking average of all $P_i$ calculated in Step I. 

\noindent Step III. To update $V$, note that under $H_1$, $Var(\delta_i) = E(\delta_i^2) = V^2+E(1/N_{Ei})$. Although we don't know absolutely whether a $\delta_i$ belongs to $H_1$, we can use posterior $P_i$ in Step II as weights:
\begin{align}\label{eq:updatev}
V^2 = \text{WAvg}(\delta_i^2;P_i) - \text{WAvg}(1/N_{Ei}; P_i)
\end{align}
where $\text{WAvg}(x_i;w_i) = \sum w_i x_i/\sum w_i $. To avoid numerical issue that $V^2$ in \eqref{eq:updatev} can take negative value, we bound $V^2$ away from $0$ by a lower bound. 

The EM algorithm starts with an initial value of $p$ and $V$, iterates through the 3 steps above until they converge. Step I is the E-step. Step II and Step III are the M-step updating $p$ and $V$.(Technically Step III is generalized M-step. We update $V$ using method of moment estimator knowing with high probability it increase the expected log-likelihood.) 

The lower bound in Step III is not purely a numerical trick. It is needed for model identification. When $V=0$, $\mu \equiv 0$ under both $H_1$ and $H_0$. We cannot distinguish $H_1$ and $H_0$, leading to an unidentifiable model. We recommend setting the lower bound $V^2 = k^2 \times \text{Avg}(1/N_E)$ and set $k$ to 2, see \citep{DengBayesAB} for explanation.
\end{document}